\documentclass[american,aps,pra,reprint, superscriptaddress]{revtex4-1}
\usepackage{amsmath,amscd,amsthm,amssymb}
\usepackage{mathrsfs}
\usepackage{graphicx}
\usepackage{caption3}
\usepackage[all]{xy}
\usepackage{rotating}
\usepackage{booktabs}
\usepackage{showlabels}
\usepackage{multirow}
\usepackage{tabularx}
\usepackage[toc,page]{appendix}
\usepackage[subfigure]{tocloft}

\usepackage[unicode=true,pdfusetitle, bookmarks=true,bookmarksnumbered=false,bookmarksopen=false, breaklinks=false,pdfborder={0 0 0},backref=false,colorlinks=false] {hyperref}
\hypersetup{ colorlinks,linkcolor=myurlcolor,citecolor=myurlcolor,urlcolor=myurlcolor}
\usepackage{graphics,epstopdf,graphicx, amsthm, amsmath, amssymb, times, braket, color, bm}
\usepackage[up]{subfigure}
\usepackage{cleveref}

\usepackage{marvosym} 

\definecolor{myurlcolor}{rgb}{0,0,0.7}

\def\be{\begin{equation}}
\def\ee{\end{equation}}
\def\bea{\begin{eqnarray*}}
\def\eea{\end{eqnarray*}}

\theoremstyle{plain}

\providecommand{\theoremname}{Theorem}

\newcommand*{\myproofname}{Proof}

\makeatother

\theoremstyle{definition}

\theoremstyle{remark}

\graphicspath{{figs/}}
\begin{document}

\title{Quantum algorithms for matrix operations and linear systems of equations}
\author{Wentao Qi}
\email{qwtmath@zju.edu.cn}
 \affiliation{School of Mathematical Sciences, Zhejiang University, Hangzhou 310027, PR~China}
 \author{Alexandr I. Zenchuk}
\email{zenchuk@itp.ac.ru}
\affiliation{Institute of Problems of Chemical Physics, Chernogolovka, Russia}
\author{Asutosh Kumar}
 \email{asutoshk.phys@gmail.com}
 \affiliation{P.G. Department of Physics, Gaya College, Magadh University, Rampur, Gaya 823001, India}
\affiliation{Harish-Chandra Research Institute, HBNI, Chhatnag Road, Jhunsi, Prayagraj 211019, India}
\affiliation{Vaidic and Modern Physics Research Centre, Bhagal Bhim, Bhinmal, Jalore 343029, India}
\author{Junde Wu}
\email{wjd@zju.edu.cn}
\affiliation{School of Mathematical Sciences, Zhejiang University, Hangzhou 310027, PR~China}

\begin{abstract}
Fundamental matrix operations and solving linear systems of equations are ubiquitous in scientific investigations. Using the ``Sender-Receiver" model, we propose quantum algorithms for matrix operations such as matrix-vector product, matrix-matrix product, the sum of two matrices, and calculation of determinant and inverse of a matrix.  We encode the matrix entries into the probability amplitudes of pure initial states of senders. After applying a proper unitary transformation to the complete quantum system, the desired result can be found in certain blocks of the receiver's density matrix.
These quantum protocols can be used as subroutines in other quantum schemes.
Furthermore, 
we present an alternative quantum algorithm for solving linear systems of equations. 
\end{abstract}

\maketitle
\section{Introduction}
Feynman and Deutsch had pointed out that it would be impossible to accurately and efficiently simulate quantum mechanical systems on a classical computer. 
They also conceived the idea that  machines ``built of quantum mechanical elements which obey quantum mechanical
laws'' \cite{Feyn, Deu} might be able to process information fundamentally more efficiently than typical classical computers. 
Such a computational power would then have applications to a multitude of problems within and outside quantum mechanics including information theory, cryptography, mathematics and statistics. 
Foreseeing such a possibility, arose the need of building ``quantum computers'' and devising ``quantum algorithms'' that could be run on quantum computers. Over the past three decades there has been a substantial growth in research on both theoretical and experimental aspects of quantum computing.
Quantum algorithms exploit principles of quantum physics, and are known to offer significant  advantages over classical counterparts for a number of problems. 
Deutsch algorithm~\cite{Deu}, a simple quantum algorithm to fast verify if a Boolean function is constant or balanced, Simon algorithm~\cite{Sim1, Sim2}, Shor's factoring algorithm~\cite{Sho1, Sho2} based on the quantum Fourier transform~\cite{QFT1, QFT2, QFT3}, and Grover's algorithm~\cite{Gro} are much celebrated.

It is possible to reduce many scientific problems to matrix formalism and to the problem of solving linear systems of equations. Fundamental matrix operations and solving linear systems of equations frequently arise on their own as well as subroutines in more complex systems, and are ubiquitous in all realms of science and engineering including machine learning and optimization. 
However, with huge data sets and large dimensions of physical systems, such tasks can be practically intractable for classical computers in terms of data processing speed, storage and time consumption. 
A quantum algorithm due to Harrow, Hassidim and Lloyd (HHL) for solving linear systems of equations~\cite{HHL} was given recently which provides an exponential speedup over the best known classical algorithms, and has led to further remarkable developments~\cite{HHL1, HHL2, HHL3}.
The HHL algorithm has been experimentally realized for some simple instances~\cite{HHL4, HHL5, HHL6, HHL7}.
Moreover, an alternative protocol was proposed for solving systems of linear algebraic equations on superconducting quantum processor of IBM Quantum Experience~\cite{SID}.
After adventing of the HHL algorithm, more efficient quantum algorithms and many substantial results for matrix operations have been obtained~\cite{Wang, ZhaoL1, Berry, ZhaoZ, Tak}. Zhao et al.~\cite{ZhaoL1}, in particular, presented a novel method for performing the elementary linear algebraic operations
on a quantum computer for complex matrices. To simulate these matrix operations, they constructed a Quantum Matrix Algebra Toolbox, a set of unitary matrices, and used the Trotter product formula together with phase estimation circuit. These results, added with quantum computation, can be used for solving many problems in machine learning and data processing.
The scalar product of arbitrary vectors has also been investigated~\cite{ZhaoZ, ZhaoL1} using an ancilla and Hadamard operator. Stolze and Zenchuk, using a ``Sender-Receiver" model, proposed an alternative scheme for the scalar product via a two-terminal quantum transmission line~\cite{Zen1}. They encoded the given vectors into the pure initial states of two different senders and obtained the result, as an element of the two-qubit receiver's density matrix, after time evolution and a proper unitary transformation. In case the model does not consider time evolution, design of the unitary transformation becomes very significant for the whole quantum computation. 

In this paper, based on the ``Sender-Receiver" model, we present quantum algorithms for matrix operations such as matrix-vector product, matrix-matrix product, sum of two matrices, and calculation of determinant and inverse of a matrix. To add further, using these quantum algorithms, 
we propose an alternative quantum scheme for solving linear systems of equations.
The proposed quantum algorithms are interesting quantum schemes for matrix operations, especially for solving linear systems of equations. In each case, the unitary operation 
$W$ that acts on the initial quantum state $\rho(0)$ of the whole system, i.e., 
$\rho = W \rho(0) W^{\dagger}$ is physically realizable. For instance, the multiple-quantum NMR technology can be used to achieve these operations.
The unitary transformation $W$, except for determinant and inverse of matrix, can be approximated,
according to the Solovay-Kitaev theorem \cite{QFT3, SK-theorem}, by the standard set of universal gates in polynomial time (see \cite{note-running-time}). 

The paper is organized as follows. In Sec. \ref{sec:sender-reciever} we recall the ``Sender-Receiver'' model, and set up the preliminary mathematical formalism. Sec. \ref{sec:quantum-algorithms} discusses the quanum algorithms for fundamental matrix operations and solving linear system of equations. Finally, we conclude in Sec. \ref{sec:conclusion}.

\begin{figure*}%
\includegraphics[width=7in]{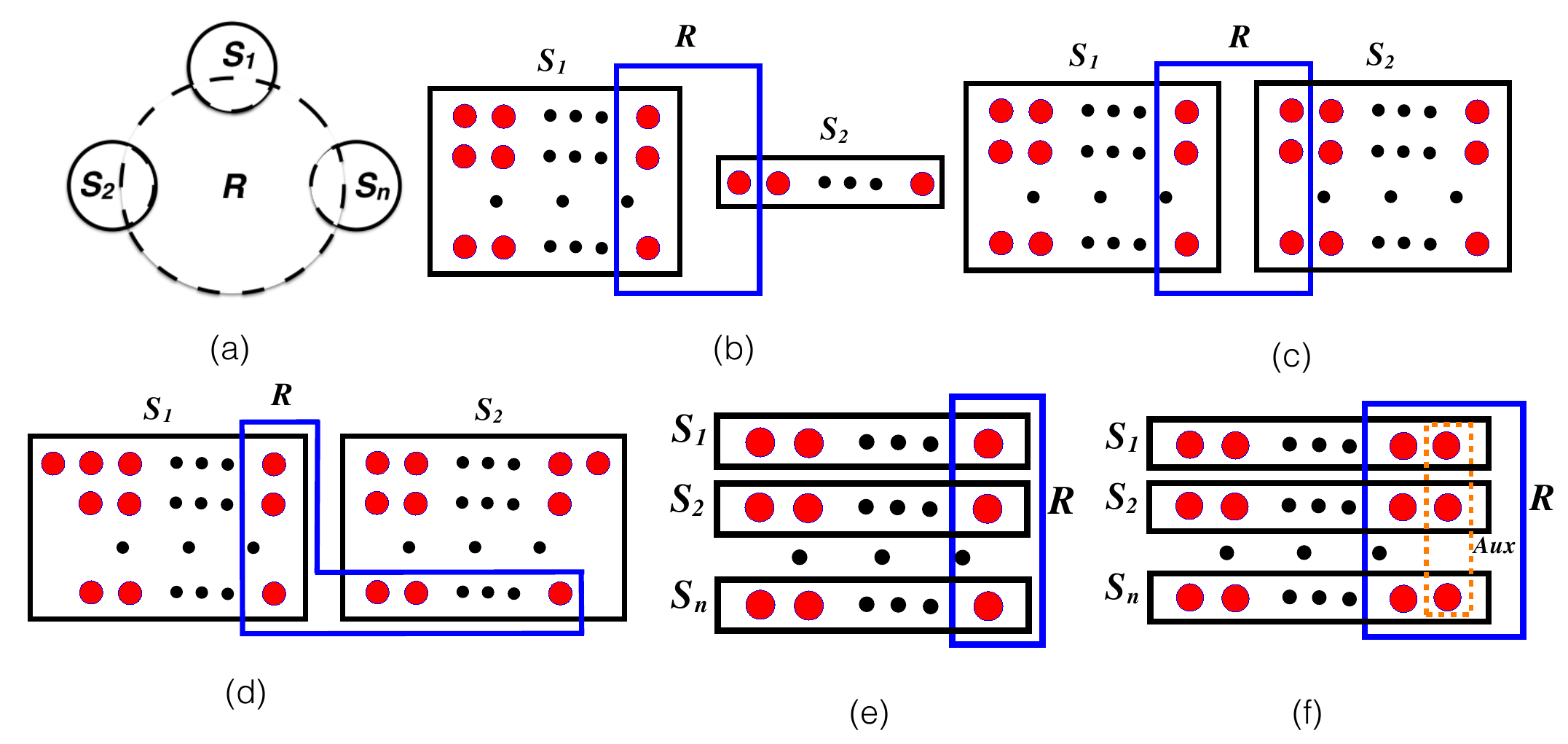}%
\caption{(Color online) The Sender (S) - Receiver (R) model for various matrix operations: (a) General schematic. (b)   Matrix-vector product. 
$S_1$ has $m$ rows of $k$-qubits and $S_2$ has only one $k$-qubit row. $R$ has $\left(m+1\right)$ qubits where the first $m$ qubits are part of $S_1$ and the last one belongs to $S_2$. 
(c) Matrix-matrix product. 
$S_1$ and $S_2$ have $m$ and $n$ rows of $k$-qubits, respectively. $R$ has $\left(m+n\right)$ qubits where the first $m$ qubits belong to $S_1$ and the remaining $n$ qubits belong to $S_2$. 
(d) Sum of two matrices. 
Both $S_1$ and $S_2$ have $m$ rows. The first row of each sender has $n+1$ qubits, and the remaining $m-1$ rows have $n$ qubits. $R$ includes $\left(m+n\right)$ qubits where the first $m$ qubits belong to the last column of $S_1$ and the rest $n$ qubits belong to the last row of $S_2$. The columns in $S_1$ are enumerated from left to right, while the columns in $S_2$ are enumerated from right to left.
(e) Determinant of an $n \times n$ square matrix. 
There are $n$ senders, and each sender $S_i$  is a single $n$-qubit row. $R$ includes the last qubit of every $S_i$.
(f) Inverse of a non-degenerate $n \times n$ square matrix. 
There are $n$ senders, and each sender $S_i$  is an $(n+1)$-qubit row. $R$ includes the last two qubits of every $S_i$.}
\label{fig:mat-op}
\end{figure*}

\section{Sender-Receiver model}\label{sec:sender-reciever}
In this model there are two or more senders and a single receiver. 
The receiver consists exclusively of some qubits of senders.
We denote the sender, the receiver and the {\it sender without receiver} respectively by $S_i$, $R$ and $S_i^{\prime}$. 
We also employ the multi-index notation \cite{FPZ_2021}. 
$X_i$ and $X_i^{\prime}$ respectively denotes all possible classical states (strings of $0$ and $1$) of the $i$th row of matrix $M$ with and without the receiver's spin (qubit). 
We conceive different ``Sender-Receiver" models for different matrix operations (see Fig. \ref{fig:mat-op}). 
For matrix-vector product, matrix-matrix product, and sum of two matrices, all entries of a particular matrix (vector) are allocated to a single sender. However, for determinant and inverse of a square matrix, we allocate elements of each row of the matrix to separate senders. Thus there are $n$ senders for an $n \times n$ matrix. Moreover, for calculating the inverse matrix, $n$ auxiliary qubits are added to the model which renders assistance for reserving $n^2$ algebraic complements. 

In our treatment, the field of matrices and vectors is complex.  
The initial states of senders are all {\it single-excitation} pure normalized states $|\psi_i\rangle$, and the elements of a matrix $M$ (a vector is also a matrix) in a particular matrix operation are encoded as the probability amplitudes in those pure states. Hence, $\left\| M \right\|_2^2 < 1$ (see \cite{note0}), where $\left\| M \right\|_2 := \sqrt{{\mbox{Tr}} (M^\dagger M )}$ is the Frobenius norm.
%
The initial state of the whole system is
\begin{equation}
\rho(0)=\rho^{\left(S_{1}\right)}(0) \otimes \cdots \otimes\rho^{\left(S_{n}\right)}(0),
\end{equation} 
where $\rho^{\left(S_{i}\right)}(0)=\left|\psi_{i}\right\rangle\left\langle\psi_{i}\right|$. $\rho(0)$ is evolved by acting an appropriate unitary transform $W$: $\rho=W \rho(0) W^{\dagger}$.
The unitary transform $W$ satisfies the commutation relation
\begin{eqnarray}\label{eqn0}
[ W,I_z ]=0,
\end{eqnarray}
where $I_z=\sum_{i=1}^{N}I_{i,z}$ is the $z$-projection operator of the total spin-momentum with every $I_{i,z}$ having eigenvalues $\pm1/2$, and $N$ is the number of spins (qubits) in the whole system. 
This commutational restriction on $W$ is an important condition for our algorithms. It ensures that $W$ is block-diagonal with respect to the number of excitations ($| 1 \rangle$s in the basis vectors of a quantum state space), and that instead of the whole state space we only need to consider some subspace constrained by certain excitation number.
If the maximal number of excitations is $n$ then $W = \text{diag} (W_0,W_1, \cdots, W_n)$.
Thus, $W$ promises that the result can be found in certain block of the receiver's density matrix $\rho^{(R)}$. 

The receiver consists of parts of the senders qubits, shown schematically by the bigger dashed circle in Fig. \ref{fig:mat-op}(a).
The receiver's state 
\begin{equation}
\rho^{(R)} := {\text {Tr}}_{S_1^{\prime} \cup \cdots \cup S_n^{\prime}} \rho
\end{equation}
is obtained after partial tracing the subsystem $S_1^{\prime} \cup \cdots \cup S_n^{\prime}$ (the whole system minus the receiver).
As mentioned earlier, each sender contains only one excitation and the unitary transform $W$ satisfies the commutation relation (\ref{eqn0}). If the maximal number of initial excitations is $n$,
the highest (by absolute value) coherence matrices~\cite{EBF, note1} included into $\rho^{(R)}$ are of the orders $\pm n$, which we denote by $\rho^{(R; \pm n)}$. Of these, the $(-n)$-order coherence matrix 
is of our significance.
Assume there are $n$ senders $S_i$ each with $r_i$ rows so that the whole system has total $r = \sum_{i=1}^n r_i$ rows.  
The receiver's state, in the multi-index notation, then can be equivalently written as
\begin{eqnarray}
\label{rhoR}
\begin{aligned}
\rho^{(R;-n)}_{N_R;M_R} &=\sum_{N^{\prime},I,J} W_{N^{\prime}N_R;I}
~\rho _{I_{S_1};J_{S_1}}^{(S_1)} \cdots\\
&\cdots \rho_{I_{S_n};J_{S_n}}^{(S_n)}~W_{J;N^{\prime}M_R}^{\dagger},
\end{aligned}
\end{eqnarray}
where $X \equiv X_1 \cdots X_r$ ($X = N^{\prime},~I,~J$), $I_{S_i} \equiv I_1 \cdots I_{r_i}$, and so on.
Aim is to find the appropriate unitary transform for every matrix operation, and present the corresponding receiver's state in a convenient form. Different strategies are adopted for different matrix operations to design the unitary transforms.

\section{Quantum algorithms}\label{sec:quantum-algorithms}
In the section we work out details of the quantum algorithms for matrix operations (see Fig. \ref{fig:mat-op}). We also illustrate quantum algorithms for obtaining determinant and solving linear system of equations.

\subsection{Matrix-vector product}\label{1}
We encode the elements of matrix $A=(a_{ij})_{m \times k}$ and vector $\mathbf{v}=(v_{j})_{k \times 1}$ as the probability amplitudes into pure normalized states 
\begin{eqnarray*}
S_1:~|\psi _1 \rangle &=& a_{00} |0\rangle+\sum_{i=1}^{m}\sum_{j=1}^{k} a_{ij} |\Psi_{i}^{(j)} \rangle~(a_{00} \neq 0), \\
S_2:~|\psi _2 \rangle &=& v_0 |0\rangle+\sum_{j=1}^{k} v_{j} |\Psi_{m+1}^{(j)} \rangle~(v_0 \neq 0),
\end{eqnarray*}
where $|\Psi_{i}^{(j)} \rangle \equiv |1\rangle_{ij} \otimes |0\rangle^{\otimes mk-1}$ is a single excited spin state of $S_1$ corresponding to the $i$th row and $j$th column, and $|\Psi_{m+1}^{(j)} \rangle \equiv |1\rangle_{j} \otimes |0\rangle^{\otimes k-1}$ is the state of $S_2$ having one excited spin at the $j$th position. There are $m+1$ rows in total.  
%
%
Since there are only two excited spins initially, and $W$ satisfies the commutation relation (\ref{eqn0}),
the highest (by absolute value) coherence matrices included into  $\rho^{(R)}$ are the $\pm2$-order coherence matrices, which we denote by $\rho^{(R;\pm 2)}$. 
%
%
We consider $M_R=\{00\cdots0\}$, and let $\rho^{(R;-2)}_{N_R;0_R}$ be the elements of the  $(-2)$-order coherence matrix. 
Let $|\cdot|$ be the sum of entries of the multi-index which equals the number of excited spins in the quantum state indicated by this index.  
Then $|N_R|=2$ and $|M_R|=0$ implies $|N'_i|=0$ in (\ref{rhoR}).  Moreover, due to the commutation relation (\ref{eqn0}), $\sum_{i=1}^{m+1} |N'_i| + |N_R| =|I|=2$ and $\sum_{i=1}^{m+1} |N'_i| + |M_R| =|J| =0 \Rightarrow |J_i|=0$.
Expression (\ref{rhoR}) therefore reduces to
\begin{eqnarray}\label{rhoR2}
&&\rho_{N_R;0_R}^{(R;-2)} =\sum_{|I_{S_1}|=1, |I_{{m+1}}|=1} W_{0_1^{\prime}\cdots 0_m^{\prime}0_{m+1}^{\prime}N_R;I_{S_1}I_{m+1}}\\\nonumber
&&
\times \rho _{I_1\cdots I_m;0_1\cdots 0_m}^{(S_1)} \rho_{I_{m+1};0_{m+1}}^{(S_2)} W_{0_1\cdots 0_{m+1} ;0_1^{\prime}\cdots 0_m^{\prime}0_{m+1}^{\prime}0_R}^{\dagger}.
\end{eqnarray}
Here $0_i$ and $0_R$ are the multi-index zeros related to the appropriate subsystems. 
Since $W_{0_1\cdots 0_{m+1} ;0_1^{\prime}\cdots 0_m^{\prime}0_{m+1}^{\prime}0_R}\equiv
W_{0_1\cdots 0_{m+1} ;0_1\cdots 0_{m+1}}$ is the only element of $0$-excitation block of $W$, we can write $W_{0_1\cdots 0_{m+1} ;0_1\cdots 0_{m+1}}=1$ 
without any loss of the generality.
Since the receiver $R$ is an $(m+1)$-qubit subsystem, its $(-2)$-order coherence matrix $\rho^{(R;-2)}$ has $\binom{m+1}{2}$ elements (number of two excitations out of $m+1$ receiver spins). But we only consider the matrix elements with index 
$$N_{R_i}=\{\underbrace{0\cdots0}_{i-1} 1
\underbrace{0 \cdots 0}_{m-i}1\}$$
with 1 at the $i$th and $(m+1)$th positions.
Eq. (\ref{rhoR2}) for these elements yields
\begin{eqnarray}
\begin{aligned}
&\rho_{N_{R_i};0_R}^{(R;-2)} =\\
&\sum_{|I_{S_1}|=1, |I_{m+1}|=1} W_{0_{S_1}^{\prime} 0_{m+1}^{\prime}N_{R_i};I_{S_1}I_{m+1}}
\rho_{I_{S_1};0_{S_1}}^{(S_1)} \rho_{I_{m+1};0_{m+1}}^{(S_2)}\\
&= \sum_{p=1}^m \sum_{|I_p|=1, |I_{m+1}|=1} W_{N_{R_i};I_pI_{m+1}} \rho_{I_p;0_{S_1}}^{(S_1)} \rho_{I_{m+1};0_{m+1}}^{(S_2)}\\
&= \sum_{p=1}^m \sum_{l,j=1}^k W_{N_{R_i};I^{(l)}_pI^{(j)}_{m+1}} \rho_{I^{(l)}_p;0_{S_1}}^{(S_1)} \rho_{I^{(j)}_{m+1};0_{m+1}}^{(S_2)},
\end{aligned}
\end{eqnarray}
where for the sake of notational convenience the $0$-parts in $W$ are omitted here, and also in later discussions unless stated otherwise. That is 
$W_{N_{R_i};I_iI_{m+1}} \equiv W_{0_1^{\prime}\cdots 0_m^{\prime}0_{m+1}^{\prime}N_{R_i};0_1\cdots0_{i-1}I_i0_{i+1}\cdots 0_mI_{m+1}}.$
%
Let
\begin{eqnarray}
W_{N_{R_i};I_p^{(l)}I_{m+1}^{(j)}}=\delta_{l,j} \delta_{i,p} \theta_1,
\end{eqnarray}
where $\theta_1=\frac{1}{\sqrt{k}}$. This involves $m$ rows of $W$, and the other rows must fulfill the unitarity.
Then with $\rho_{I_i^{(j)};0_{S_1}}^{(S_1)}=a_{ij}a_{00}^*$, $\rho_{I_{m+1}^{(j)};0_{m+1}}^{(S_2)}=v_jv_0^*$ and $\alpha=\theta_1a_{00}^*v_0^*$, we have
$\rho_{N_{R_i};0_R}^{(R;-2)}=\alpha\sum_{j=1}^{k}a_{ij}v_{j}$.
%
Hence, the matrix-vector product is
\begin{equation}
A\mathbf{v}=\frac{1}{\alpha} \left(
\rho_{N_{R_i};0_R}^{(R;-2)} \right)_{m \times 1}.
\end{equation}

\subsection{Matrix-matrix product }\label{2}
We consider the matrix product $AB$ of two matrices $A=(a_{ij})_{m \times k}$ and $B=(b_{ij})_{k \times n}$. See Fig. \ref{fig:mat-op}(c). The initial states of two senders are
\begin{eqnarray*}
S_1:~|\psi _1 \rangle &=& a_{00} |0\rangle+\sum_{i=1}^{m}\sum_{j=1}^{k} a_{ij} |\Psi_{i}^{(j)} \rangle~(a_{00} \neq 0), \\
S_2:~|\psi _2 \rangle &=& b_{00} |0\rangle+\sum_{i=1}^{n}\sum_{j=1}^{k} b_{ji} |\Psi_{m+i}^{(j)} \rangle~(b_{00} \neq 0).
\end{eqnarray*}
Here $|\Psi_{i}^{(j)} \rangle \equiv |1\rangle_{ij} \otimes |0\rangle^{\otimes mk-1}$ and $|\Psi_{m+i}^{(j)} \rangle \equiv |1\rangle_{ij} \otimes |0\rangle^{\otimes nk-1}$  are the single-excitation states of $S_1$ and $S_2$ respectively. Note that the transpose of $B$ is encoded into the quantum state $|\psi_2\rangle$ of $S_2$. That is, the probability amplitude of $|\Psi_{m+i}^{(j)} \rangle$ is $b_{ji}$ (rather than $b_{ij}$) to satisfy the multiplication rule. 
In the following, we carry out similar treatment as in matrix-vector product. We consider the multi-index $N_R$ in which the entries corresponding to the $i$th and the $(m+j)$th spin, respectively, in the first and second columns of $R$ are $1$, and others are $0$. That is,
\begin{eqnarray}\label{NRij}
N_{R_{ij}}=\{\underbrace{0\cdots0}_{i-1}1\underbrace{0\cdots0}_{m-i+j-1}1\underbrace{0\cdots0}_{n-j}\}.
\end{eqnarray}
Then we have
\begin{eqnarray}
\begin{aligned}
\rho_{N_{R_{ij}};0_R}^{(R;-2)}&=\sum_{p=1}^m\sum_{q=1}^n
\sum_{|I_p|=|I_{m+q}|=1} W_{N_{R_{ij}};I_pI_{m+q}}\\
&\times\rho_{I_p;0_{S_1}}^{(S_1)}\rho_{I_{m+q};0_{S_2}}^{(S_2)}\\
&=\sum_{p=1}^m\sum_{q=1}^n \sum_{l,h=1}^{k}W_{N_{R_{ij}};I_p^{(l)}I_{m+q}^{(h)}}\rho_{I_p^{(l)};0_{S_1}}^{(S_1)}\rho_{I_{m+q}^{(h)};0_{S_2}}^{(S_2)}.
\end{aligned}
\end{eqnarray}

Now let
\begin{eqnarray}
W_{N_{R_{ij}};I_p^{(l)}I_{m+q}^{(h)}}=\delta_{l,h} \delta_{i,p}\delta_{j,q}\theta_1,
\end{eqnarray}
and with $\rho_{I_i^{(l)};0_{S_1}}^{(S_1)}=a_{il}a_{00}^*$, $\rho_{I_{m+j}^{(l)};0_{S_2}}^{(S_2)}=b_{lj}b_{00}^*$ and $\beta=\theta_1a_{00}^*b_{00}^*$, we have
$\rho_{N_{R_{ij}};0_R}^{(R;-2)} = \beta\sum_{l=1}^{k}a_{il}b_{lj}$.
%
Thus, the matrix product is
\begin{equation}
AB=\frac{1}{\beta} \,\left( \rho_{N_{R_{ij}};0_R}^{(R;-2)} \right)_{m \times n}.
\end{equation}

\subsection{Sum of two matrices}\label{2_2}
Consider two $m \times n$ matrices $C$ and $D$. The first row of each sender has $n+1$ elements, while the remaining rows have $n$ elements.
The initial states of two subsystems read
\begin{eqnarray*}
S_1:~|\psi _1 \rangle &=& c_{00} |0\rangle+\sum_{i=1}^{m}\sum_{j=1}^{n} c_{ij} |\Lambda_{i}^{(j)} \rangle +\lambda |\Lambda_1^{(0)}\rangle, \\
S_2:~|\psi _2 \rangle &=& d_{00} |0\rangle+\sum_{i=1}^{m}\sum_{j=1}^{n} d_{ij} |\Lambda_{m+i}^{(j)} \rangle + \lambda |\Lambda_{m+1}^{(0)}\rangle,
\end{eqnarray*}
where $c_{00}\neq 0, \;d_{00}\neq 0,\; \lambda\neq 0$. The $|\Lambda_i^{(j)}\rangle$ and $|\Lambda_{m+i}^{(j)}\rangle$ are the product states of single-excitation $|1 \rangle$ and $|0 \rangle^{\otimes mn}$. The receiver consists of the last column of $S_1$ and the last row of  $S_2$ as seen in Fig. \ref{fig:mat-op}(d).

Here elements of the $(-2)$-order coherence matrix of the receiver density matrix $\rho^{(R)}$ can be written as $\rho^{(R;-2)}_{N_R,0_R} =\sum_{I_{S_1}I_{S_2}} W_{0'_{S_1}0'_{S_2}N_R;I_{S_1}I_{S_2}} \rho^{(S_1)}_{I_{S_1},0_{S_1}} \rho^{(S_2)}_{I_{S_2},0_{S_2}}$.
We choose elements of the receiver density matrix $\rho^{(R)}$ marked by the index $N_{R_{ij}}$ in Eq. (\ref{NRij}) corresponding to the $i$th excited spin of the column and $j$th excited spin of the row of the receiver. Hence, 
\begin{eqnarray}
\rho^{(R;-2)}_{N_{R_{ij}};0_R} = \sum_{p,q,l,h} W_{N_{R_{ij}};I_{p}^{(l)} I_{m+q}^{(h)}} \rho^{(S_1)}_{I_{p}^{(l)};0_{S_1}} \rho^{(S_2)}_{I_{m+q}^{(h)};0_{S_2}}.
\end{eqnarray}
There are $mn$ such elements which coincides with the total number of elements in either matrix $C$ or $D$. Therefore, the elements $\rho^{(R;-2)}_{N_{R_{ij}},0_R}$ can be used to store the sum $C+D$. Let
\begin{eqnarray}
W_{N_{R_{ij}};I_{p}^{(l)} I_{m+q}^{(h)}} 
= (\delta_{ip} \delta_{jl} \delta_{q1}\delta_{h0} +
\delta_{iq} \delta_{jh} \delta_{p1}\delta_{l0}) \theta_2,
\end{eqnarray}
where $\theta_2=\frac{1}{\sqrt{2}}$.
Then with $\omega=  \theta_2 c_{00}^* d_{00}^*$,
\begin{eqnarray}
\rho^{(R;-2)}_{N_{R_{ij}};0_R} &&= \theta_2 (\rho^{(S_1)}_{I_i^{(j)};0_{S_1}} \rho^{(S_2)}_{I_{m+1}^{(0)};0_{S_2}} + \rho^{(S_2)}_{I_{m+i}^{(j)};0_{S_2}} \rho^{(S_1)}_{I_1^{(0)};0_{S_1}}\\\nonumber
&&=\omega \lambda (c_{ij}+ d_{ij})
\end{eqnarray}
is proportional to the sum of two proper elements of $C$ and $D$, 
where $I^{(0)}_1$ and $I^{(0)}_{m+1}$ are extra elements of the first row of $S_1$ and $S_2$, respectively. Hence, 
\begin{eqnarray}
C+D=\frac{1}{\omega \lambda}\left(\rho^{(R;-2)}_{N_{R_{ij}},0_R}\right)_{m \times n}.
\end{eqnarray}

\subsection{Determinant}\label{3}
To compute the determinant of a square matrix $E = (e_{ij})_{n \times n}$, we encode every $i$th row of $E$ into the pure initial state of sender $S_i$
\begin{eqnarray*}
|\psi_i \rangle=e_{i0}|0\rangle+\sum_{j=1}^{n}e_{ij}| \Phi_{i}^{(j)} \rangle~(e_{i0}\neq 0),
\end{eqnarray*}
where $|\Phi_i^{(j)} \rangle \equiv |1\rangle_j \otimes |0 \rangle^{\otimes n-1}$ represents the state of $S_i$ with only one excited spin at the $j$th position. Here, contrary to the previous investigations, there are $n$ senders and one receiver. Consequently, a slight modification in the treatment is in order.

We only consider $N_R=\{ 11\cdots1 \}$ and $M_R=\{ 00\cdots0 \}$, so that $\rho_{N_R;M_R}^{(R)} \equiv \rho_{1_R;0_R}^{(R)}$ is exactly the unique element of the receiver's $(-n)$-order coherence matrix $\rho^{(R;-n)}$,
\begin{eqnarray}
\rho_{1_R;0_R}^{(R;-n)}=\sum_{|I_1|=1}\cdots\sum_{|I_n|=1}W_{1_R;I_1\cdots I_{n}}\rho_{I_1;0_1 }^{(S_1)}\cdots\rho_{I_{n};0_n}^{(S_n)}.
\end{eqnarray}

Let
\begin{eqnarray}
W_{1_R;I_1^{(i_1)}\cdots I_{n}^{(i_n)}}=\epsilon_{i_1i_2\cdots i_n}\theta_3,
\end{eqnarray}
where $\theta_3=\frac{1}{\sqrt{n!}}$ and $\epsilon_{i_1i_2\cdots i_n}$ is the permutation symbol which equals $1 (-1)$ if the inversion number of permutation $i_1i_2\cdots i_n$ is even (odd), and $0$ if at least two indices coincide.
Then with $\rho_{I_l^{(i_l)};0_l}^{(S_l)}=e_{li_l}e_{l0}^*$ and $\gamma=\theta_3\prod\limits_{i = 1}^n {{e_{i0}}^*}$,
\begin{eqnarray}
\begin{aligned}
\rho_{1_R;0_R}^{(R;-n)}&=\theta_3\sum_{i_1i_2\cdots i_n}
\epsilon_{i_1i_2\cdots i_n}
\rho_{I_1^{(i_1)};0_1}^{(S_1)}\cdots\rho_{I_{n}^{(i_n)};0_n}^{(S_n)}\\
&=\gamma\sum_{i_1i_2\cdots i_n}
\epsilon_{i_1i_2\cdots i_n}
e_{1i_1}e_{2i_2}\cdots e_{ni_n}.
\end{aligned}
\end{eqnarray}
Hence, the determinant of matrix $E$ reads
\begin{equation}
\det(E)=\frac{1}{\gamma} \;\rho_{1_R;0_R}^{(R;-n)}.
\end{equation}

{\it Illustration for obtaining determinant.}--
For the square matrix
$E=\left(
\begin{array}{cc}
3/4 & 1/4 \\
1/4 & 3/4
\end{array}
\right)$, 
the pure initial states of two senders can be written as 
\begin{eqnarray*}
S_1:~|\psi_1 \rangle &=& \frac{\sqrt{6}}{4}|00\rangle+\frac{3}{4}|10\rangle+\frac{1}{4}|01\rangle, \\
S_2:~|\psi_2 \rangle &=& \frac{\sqrt{6}}{4}|00\rangle+\frac{1}{4}|10\rangle+\frac{3}{4}|01\rangle,
\end{eqnarray*}
where $e_{10}=e_{20}=\frac{\sqrt{6}}{4}$, and the initial state of the whole system is  
\begin{eqnarray*}
\rho(0)&=&\rho^{\left(S_{1}\right)}(0) \otimes \rho^{\left(S_{2}\right)}(0) = \left|\psi_{1}\psi_{2}\right\rangle\left\langle\psi_{1}\psi_2\right| \\
&=&\cdots+\frac{3}{4}\cdot\frac{3}{4}\cdot\frac{\sqrt{6}}{4}\cdot\frac{\sqrt{6}}{4}|1001\rangle\langle0000| \\
&+&\frac{1}{4}\cdot\frac{1}{4}\cdot\frac{\sqrt{6}}{4}\cdot\frac{\sqrt{6}}{4}|0110\rangle\langle0000|+\cdots.
\end{eqnarray*}
We need to design the unitary transformation $W$ to find one entry in $W \rho(0) W^{\dagger}$ showing the determinant multiplied by a constant, that is, 
$$\frac{3}{4}\cdot\frac{3}{4}\cdot\frac{\sqrt{6}}{4}\cdot\frac{\sqrt{6}}{4}-\frac{1}{4}\cdot\frac{1}{4}\cdot\frac{\sqrt{6}}{4}\cdot\frac{\sqrt{6}}{4}.$$
Let $W_{\cdot;1001}=\frac{1}{\sqrt{2}}$ and $W_{\cdot;0110}=-\frac{1}{\sqrt{2}}$, where $\sqrt{2}$ is designed to ensure that $W$ is a unitary matrix. The commutation relation $[ W,I_z ]=0$ shows that the $W$ is block-diagonal with respect to the number of excitations. Since the 4-qubit system contains no more than four excitations, the unitary transformation $W$ includes 5 blocks: $W={\text {diag}}(W_0,W_1,W_2,W_3,W_4)$. We let $W_0=W_4=1$ and $W_1=W_3=I$. The block of major concern here is $W_2$ written in the basis $\{ \text{permut}~ |0^{\otimes 2} 1^{\otimes 2} \rangle \} = \{ |0011\rangle, |0101\rangle, \cdots, |1100\rangle \}$:
$$
W_2=\left( 
\begin{array}{cccccc}
0 & 0 & \frac{-1}{\sqrt{2}} & \frac{1}{\sqrt{2}} & 0 & 0 \\ 
0 & 0 & \frac{1}{\sqrt{2}} & \frac{1}{\sqrt{2}} & 0 & 0 \\ 
1 & 0 & 0 & 0 & 0 & 0 \\ 
0 & 1 & 0 & 0 & 0 & 0 \\ 
0 & 0 & 0 & 0 & 1 & 0 \\ 
0 & 0 & 0 & 0 & 0 & 1%
\end{array}%
\right). 
$$

After partial tracing the 1st and the 3rd qubits, we have 
\begin{eqnarray*}
\rho^{(R)}&=& \sum_{x,y=0}^1(\langle xy| \otimes I)W\rho(0) W^{\dagger} (|xy \rangle \otimes I)\\
&=& \left( 
\begin{array}{cccc}
\frac{3}{8} & \frac{5\sqrt{3}}{128}+\frac{27\sqrt{6}}{256} & \frac{15\sqrt{6}%
}{256} & \frac{3\sqrt{2}}{32} \\ 
\frac{5\sqrt{3}}{128}+\frac{27\sqrt{6}}{256} & \frac{113}{256} & \frac{27}{%
	256} & \frac{3\sqrt{3}}{32} \\ 
\frac{15\sqrt{6}}{256} & \frac{27}{256} & \frac{15}{256} & \frac{\sqrt{3}}{32%
} \\ 
\frac{3\sqrt{2}}{32} & \frac{3\sqrt{3}}{32} & \frac{\sqrt{3}}{32} & \frac{1}{%
	8}%
\end{array}%
\right).
\end{eqnarray*}
Hence, $\rho_{11;00}^{(R;-2)}=\frac{3\sqrt{2}}{32}$. Also, $e_{10}=e_{20}=\frac{\sqrt{6}}{4}$ and $\theta_3=\frac{1}{\sqrt{2}}$ yield $\gamma=\theta_3e_{10}e_{20}=\frac{3\sqrt{2}}{16}$.
Therefore, the determinant is 
$\det(E)=\frac{1}{\gamma} \;\rho_{11;00}^{(R;-2)}=\frac{1}{2}.$

\subsection{Inverse of matrix}\label{4}
Like determinant, to compute the inverse of a non-degenerate square matrix $E = (e_{ij})_{n \times n}$, we encode every $i$th row of $E$ into the pure state of sender $S_i$. Morever, unlike previous ``Sender-Receiver'' models, we add the auxiliary module called $Aux$ to the senders (the last qubits of the senders). These  auxiliary qubits appear in $R$ which includes two last spins of each sender. See Fig. \ref{fig:mat-op}(f). The normalized initial state of each sender $S_i$ is given by
\begin{eqnarray*}
|\psi_i \rangle&=\hat e_{i0}|0\rangle+\sum_{j=1}^{n}e_{ij}| \Xi_{i}^{(j)}\rangle+\sigma| \Xi_{i}^{(n+1)} \rangle,
\end{eqnarray*}
where $\hat e_{i0} \neq 0,~\sigma\neq 0$, $\hat e_{ij}$ are the elements of the non-degenerate matrix $E$ whose inverse is intended, and $| \Xi_{i}^{(j)} \rangle \equiv |1\rangle_j \otimes |0 \rangle^{\otimes n}$ is the state of sender $S_i$ corresponding to the $j$th excited spin (qubit). In particular, $| \Xi_{i}^{(n+1)} \rangle$ is the state associated with the excited spin of $S_i$ included in the $Aux$ module.
We have already obtained the determinant. To find the inverse $E^{-1}=E^*/ \text {det}(E)$, 
where $E^*$ is the adjoint matrix, our goal is to find every algebraic complement $E_{ij}$ of  $e_{ij}$ (the matrix element of $E$).

We consider the $(-n)$-order coherence matrix with elements $
\rho^{(R;-n)}_{N_R;0_R}$
and select $n^2$ elements as follows. The receiver $R$ can be viewed as a two-column subsystem. Let the entries of the multi-index $N_R$ assotiated with the first column (the $n$th spins of each sender) be 1 except the $i$th entry which is 0. Similarly, let the entries of the multi-index $N_R$ assotiated with the second column (i.e., the entries corresponding to the module $Aux$) be 0 except the $j$th one which is 1. We denote such multi-index by $\hat N_{R_{ij}}$, $i,j=1,\dots,n$. Thus, there are $n^2$ different elements in selected part of $\rho^{(R;-n)}$. These are used for storing the elements of the inverse matrix $E^{-1}$.

The elements $\rho^{(R;-n)}_{\hat N_{R_{ij}};0_R}$ are
\begin{eqnarray}
\begin{aligned}
\rho_{\hat N_{R_{ij}};0_R}^{(R;-n)}&=\sum_{N_1^{\prime},\cdots,N_n^{\prime},I,J} W_{N_1^{\prime}\cdots N_n^{\prime}\hat N_{R_{ij}};I} \rho _{I_1;J_1}^{(S_1)} \cdots\\
&\cdots \rho_{I_{n};J_{n}}^{(S_n)}W_{J;N_1^{\prime}\cdots N_n^{\prime}0_R}^{\dagger}\\
&=\sum_{|I_1|=1}\cdots\sum_{|I_n|=1}W_{\hat N_{R_{ij}};I_1\cdots I_{n}}\rho_{I_1;0_1}^{(S_1)}\cdots\rho_{I_{n};0_n}^{(S_n)}.
\end{aligned}
\end{eqnarray}

Let
\begin{eqnarray}\label{eqn2}
\begin{aligned}
&\quad W_{\hat N_{R_{ij}};I_1^{(l_1)}\cdots I_{i-1}^{(l_{i-1})}I_i^{({n+1})}I_{i+1}^{(l_{i+1})}\cdots I_{n}^{(l_n)}}\\
&=(-1)^{i+j}
\epsilon_{j;l_1\cdots l_{i-1}l_{i+1}\cdots l_n}
\theta_4,
\end{aligned}
\end{eqnarray}
where $\theta_4=\frac{1}{\sqrt{(n-1)!}}$ and
$\epsilon_{j;l_1\cdots l_{i-1}l_{i+1}\cdots l_n}$ is $+1 ~(-1)$
if $\{l_1\cdots l_{i-1}l_{i+1}\cdots l_n\}$
is an even (odd) permutation of
$\{1,\cdots, j-1,j+1, \cdots n\}$, and $0$ otherwise.
%
Thus, we obtain
\begin{eqnarray}
\begin{aligned}
&\quad \rho_{\hat N_{R_{ij}};0_R}^{(R;-n)}\\
&=(-1)^{i+j}\theta_4\sum_{l_1\cdots l_{i-1}l_{i+1}\cdots l_n}
\epsilon_{j;l_1\cdots l_{i-1}l_{i+1}\cdots l_n}
\rho_{I_1^{(l_1)};0_1}^{(S_1)}\cdots\\
&\cdots\rho_{I_{i-1}^{(l_{i-1})};0_{i-1}}^{(S_{i-1})}\rho_{I_i^{(n+1)};0_i}^{(S_i)}\rho_{I_{i+1}^{(l_{i+1})};0_{i+1}}^{(S_{i+1})}\cdots \rho_{I_{n}^{(l_n)};0_{n}}^{(S_n)}\\
&=(-1)^{i+j}\mu\sum_{l_1\cdots l_{i-1}l_{i+1}\cdots l_n}
\epsilon_{j;l_1\cdots l_{i-1}l_{i+1}\cdots l_n}
\\
&\times e_{1l_1}\cdots e_{i-1,l_{i-1}}e_{i+1,l_{i+1}}\cdots e_{ni_n},
\end{aligned}
\end{eqnarray}
where $\mu=\theta_4\sigma\prod\limits_{i = 1}^n {{\hat e_{i0}^*}}$. Thus, the algebraic complement of $e_{ij}$ is
\begin{eqnarray}
E_{ij}=\frac{1}{\mu} \, \rho_{\hat N_{R_{ij}};0_R}^{(R;-n)}.
\end{eqnarray}

We can also find the determinant of $E$ from $\rho^{(R)}$ by continuing to design the unitary transformation $W$. Let $\hat N_{R}=\{ 1010\cdots 10 \}$ be the element $\rho^{(R;-n)}_{\hat N_R;0_R}$ which corresponds to  the states $|0\rangle$ of all spins from $Aux$ and states $|1\rangle$ of all spins from the first column of $R$. Let the elements of $W$ satisfy the constraint:
\begin{eqnarray}
W_{\hat N_R;I_1^{(i_1)}\cdots I_{n}^{(i_n)}}=
\epsilon_{i_1i_2\cdots i_n}\theta_3
\end{eqnarray}
in addition to the constraint (\ref{eqn2}).
Hence,
\begin{eqnarray}
\begin{aligned}
\rho_{\hat N_{R};0_{R}}^{(R;-n)}&=\theta_3\sum_{i_1i_2\cdots i_n}
\epsilon_{i_1i_2\cdots i_n}
\rho_{I_1^{(i_1)};0_1}^{(S_1)}\cdots\rho_{I_{n}^{(i_n)};0_n}^{(S_n)}\\
&=\hat \gamma\sum_{i_1i_2\cdots i_n}
\epsilon_{i_1i_2\cdots i_n}
e_{1i_1}e_{2i_2}\cdots e_{ni_n}
\end{aligned}
\end{eqnarray}
where $\hat \gamma=\theta_3\prod\limits_{i = 1}^n {{\hat e_{i0}^*}}$.
The determinant of $E$ is
$\det(E)=\frac{1}{\hat \gamma} \;\rho_{\hat N_{R};0_{R}}^{(R;-n)}$.
Hence, the inverse of $E$ is
\begin{eqnarray}
E^{-1}=\frac{\hat \gamma}{\rho_{\hat N_{R};0_{R}}^{(R;-n)}}E^*
\end{eqnarray}
with $E^*(j,i)=\frac{1}{\mu} \, \rho_{\hat N_{R_{ij}};0_R}^{(R;-n)}$. Specifically,
\begin{eqnarray}\label{eqn3}
E^{-1}(j,i)=\frac{\rho_{\hat N_{R_{ij}};0_R}^{(R;-n)}}{\sigma\sqrt{n}\rho_{\hat N_{R};0_{R}}^{(R;-n)}}.
\end{eqnarray}

See Table \ref{table:quant-algo} for a summary of quantum algorithms for matrix operations.

\subsection{Linear system of equations}
We apply quantum algorithms investigated in this paper to solve linear system of equations, $E\mathbf{x}=\mathbf{b}$, where $E = (e_{ij})_{n \times n}$ is a non-degenerate matrix and $\mathbf{b}=(b_1,b_2,\cdots,b_n)^T$ is a unit vector.
To obtain the solution $\mathbf{x}=E^{-1}\mathbf{b}$, we first construct the unitary transformation $V$ satisfying the condition (\ref{eqn0}) by encoding $\mathbf{b}$ into $V$, and then act it on the receiver's state $\rho^{(R)}$ of the inverse matrix operation: $\xi^{(R)} = V \rho^{(R)}  V^{\dagger}$.
%
Let
\begin{eqnarray}
V_{L_{R_{j}};\hat N_{R_{ij}}}=b_{i},
\end{eqnarray}
where $L_{R_{j}}=\hat N_{R_{jj}} =\{\underbrace{10\cdots10}_{2(j-1)} 01
\underbrace{10 \cdots 10}_{2(n-j)}\}$,
and other entries need to fulfill the unitarity of $V$. 
%
%
Now with $V_{0_{R};0_{R}}^{\dagger}=1$ and using Eq. (\ref{eqn3}), we have
\begin{eqnarray}
\begin{aligned}
\xi_{L_{R_{j}};0_R}^{(R;-n)} &= \sum_{i=1}^{n}V_{L_{R_{j}};\hat N_{R_{ij}}}\rho_{\hat N_{R_{ij}};0_R}^{(R;-n)}V_{0_{R};0_{R}}^{\dagger}\\
&=\sigma\sqrt{n}\rho_{\hat N_{R};0_{R}}^{(R;-n)}\sum_{i=1}^{n}E^{-1}(j,i)b_i\\
&= \sigma\sqrt{n}\rho_{\hat N_{R};0_{R}}^{(R;-n)}x_j
\end{aligned}
\end{eqnarray}

Hence, the solution is
\begin{eqnarray}\label{xi}
\mathbf{x}=\frac{1}{\sigma\sqrt{n}\rho_{\hat N_{R};0_{R}}^{(R;-n)}}
\left(
\xi_{L_{R_{i}};0_R}^{(R;-n)}
\right)_{n \times 1}.
\end{eqnarray}

{\it Illustration for solving linear system of equations.}--
We choose
\begin{eqnarray}
E=\left(
\begin{array}{cc}
3/4 & 1/4 \\
1/4 & 3/4
\end{array}
\right)~\text{and}~
\mathbf{b}=\frac{1}{\sqrt{2}}\left(
\begin{array}{c}
1\\
1
\end{array}
\right)
\end{eqnarray}
in $E\mathbf{x}=\mathbf{b}$ so that
\begin{eqnarray}
\mathbf{x}=\frac{1}{\sqrt{2}}\left(
\begin{array}{c}
1\\
1
\end{array}
\right).
\end{eqnarray}

We fix the parameters $\hat e_{10}=\hat e_{20} = \frac{1}{2}$ and $\sigma=\frac{1}{2\sqrt{2}}$, so that the pure initial states of two senders are
\begin{eqnarray*}
S_1:~|\psi_1 \rangle &=& \frac{1}{2}|000\rangle+\frac{3}{4}|100\rangle+\frac{1}{4}|010\rangle+\frac{1}{2\sqrt{2}}|001\rangle, \\
S_2:~|\psi_2 \rangle &=& \frac{1}{2}|000\rangle+\frac{1}{4}|100\rangle+\frac{3}{4}|010\rangle+\frac{1}{2\sqrt{2}}|001\rangle.
\end{eqnarray*}

The unitary transformation $W$ is block-diagonal, $W={\text {diag}}(1,W_1,W_2,\cdots,W_6)$
and we choose the two-excitation block $W_2$ in the basis $\{ \text{permut}~ |0^{\otimes 4} 1^{\otimes 2} \rangle \} = \{ |000011\rangle, |000101\rangle, \cdots, |110000\rangle \}$ of the whole system as:
\begin{eqnarray}
\begin{aligned}
&\quad W_2=\\
&\left(
\begin{array}{ccccccccccccccc}
0 & 0 & 0 & 0 & 0 & -1 & 0 & 0 & 0 & 0 & 0 & 0 & 0 & 0 & 0 \\
0 & 1 & 0 & 0 & 0 & 0 & 0 & 0 & 0 & 0 & 0 & 0 & 0 & 0 & 0 \\
0 & 0 & 1 & 0 & 0 & 0 & 0 & 0 & 0 & 0 & 0 & 0 & 0 & 0 & 0 \\
0 & 0 & 0 & 1 & 0 & 0 & 0 & 0 & 0 & 0 & 0 & 0 & 0 & 0 & 0 \\
0 & 0 & 0 & 0 & 1 & 0 & 0 & 0 & 0 & 0 & 0 & 0 & 0 & 0 & 0 \\
1 & 0 & 0 & 0 & 0 & 0 & 0 & 0 & 0 & 0 & 0 & 0 & 0 & 0 & 0 \\
0 & 0 & 0 & 0 & 0 & 0 & 0 & 0 & 0 & 0 & 1 & 0 & 0 & 0 & 0 \\
0 & 0 & 0 & 0 & 0 & 0 & 0 & 0 & \frac{-1}{\sqrt{2}} & 0 & 0 & \frac{1}{\sqrt{%
		2}} & 0 & 0 & 0 \\
0 & 0 & 0 & 0 & 0 & 0 & 0 & 1 & 0 & 0 & 0 & 0 & 0 & 0 & 0 \\
0 & 0 & 0 & 0 & 0 & 0 & -1 & 0 & 0 & 0 & 0 & 0 & 0 & 0 & 0 \\
0 & 0 & 0 & 0 & 0 & 0 & 0 & 0 & \frac{1}{\sqrt{2}} & 0 & 0 & \frac{1}{\sqrt{2%
}} & 0 & 0 & 0 \\
0 & 0 & 0 & 0 & 0 & 0 & 0 & 0 & 0 & 1 & 0 & 0 & 0 & 0 & 0 \\
0 & 0 & 0 & 0 & 0 & 0 & 0 & 0 & 0 & 0 & 0 & 0 & 1 & 0 & 0 \\
0 & 0 & 0 & 0 & 0 & 0 & 0 & 0 & 0 & 0 & 0 & 0 & 0 & 1 & 0 \\
0 & 0 & 0 & 0 & 0 & 0 & 0 & 0 & 0 & 0 & 0 & 0 & 0 & 0 & 1%
\end{array}%
\right).
\end{aligned}
\end{eqnarray}

Also, the unitary transformation $V={\text {diag}}(1,V_1,V_2,V_3,V_4)$ is  block-diagonal where $V_2$ in the basis $\{ \text{permut}~ |0^{\otimes 2} 1^{\otimes 2} \rangle \} = \{ |0011\rangle, |0101\rangle, \cdots, |1100\rangle \}$ of the receiver is
\begin{eqnarray}
V_2=\left(
\begin{array}{cccccc}
\frac{-1}{\sqrt{2}} & 0 & 0 & \frac{1}{\sqrt{2}} & 0 & 0 \\
0 & 1 & 0 & 0 & 0 & 0 \\
0 & 0 & \frac{1}{\sqrt{2}} & 0 & 0 & \frac{1}{\sqrt{2}} \\
\frac{1}{\sqrt{2}} & 0 & 0 & \frac{1}{\sqrt{2}} & 0 & 0 \\
0 & 0 & 0 & 0 & 1 & 0 \\
0 & 0 & \frac{1}{\sqrt{2}} & 0 & 0 & \frac{-1}{\sqrt{2}}%
\end{array}%
\right).
\end{eqnarray}

With unitary transformations $W$ and $V$, we can find the solution of linear system of equations.
We can choose $W_i=I$ and $V_j=I$ for $i,j \neq2$.

\begin{table*}[!htbp]
\begin{tabular}{|l|l|l|l|} 
\hline
Matrix operation      & Senders' states                                                                                                               & Unitary transfrom & Receiver's state \\ \hline

Matrix-vector product \\ $A=(a_{ij})_{m \times k}$, $\mathbf{v}=(v_{j})_{k \times 1}$ & \begin{tabular}[c]{@{}l@{}}$S_1:~|\psi_1\rangle = a_{00} |0\rangle $\\$+ \sum_{i=1}^{m}\sum_{j=1}^{k} a_{ij} |\Psi_{i}^{(j)} \rangle$ \\~\\ $S_2:~| \psi _2 \rangle =v_0 |0\rangle$\\$+\sum_{j=1}^{k} v_{j} |\Psi_{m+1}^{(j)} \rangle$\end{tabular} & $W_{N_{R_i};I_p^{(l)}I_{m+1}^{(j)}}=\delta_{l,j} \delta_{i,p} \theta_1$     & \begin{tabular}[c]{@{}l@{}}$A\mathbf{v}=\frac{1}{\alpha} \left(
\rho_{N_{R_i};0_R}^{(R;-2)} \right)_{m \times 1}$ \\~\\ $\alpha=\theta_1a_{00}^*v_0^*$\end{tabular}      \\ \hline

Matrix-matrix product \\ $A=(a_{ij})_{m \times k}$, $B=(b_{ij})_{k \times n}$ & \begin{tabular}[c]{@{}l@{}}$S_1:~|\psi _1 \rangle = a_{00} |0\rangle$\\$+\sum_{i=1}^{m}\sum_{j=1}^{k} a_{ij} |\Psi_{i}^{(j)} \rangle$ \\~\\
$S_2:~|\psi _2 \rangle = b_{00} |0\rangle$\\$+\sum_{i=1}^{n}\sum_{j=1}^{k} b_{ji} |\Psi_{m+i}^{(j)} \rangle$\end{tabular}                   & $W_{N_{R_{ij}};I_p^{(l)}I_{m+q}^{(h)}}=\delta_{l,h} \delta_{i,p}\delta_{j,q}\theta_1$   & \begin{tabular}[c]{@{}l@{}}$AB=\frac{1}{\beta} \,\left( \rho_{N_{R_{ij}};0_R}^{(R;-2)} \right)_{m \times n}$  \\~\\ $\beta=\theta_1a_{00}^*b_{00}^*$\end{tabular}        \\ \hline

Sum of matrices  \\ $C=(c_{ij})_{m \times n}$, $D=(d_{ij})_{m \times n}$     & \begin{tabular}[c]{@{}l@{}}$S_1:~|\psi _1 \rangle = c_{00} |0\rangle +\lambda |\Lambda_1^{(0)}\rangle $\\$+\sum_{i=1}^{m}\sum_{j=1}^{n} c_{ij} |\Lambda_{i}^{(j)} \rangle $ \\~\\
$S_2:~|\psi _2 \rangle = d_{00} |0\rangle+ \lambda |\Lambda_{m+1}^{(0)}\rangle$\\$+\sum_{i=1}^{m}\sum_{j=1}^{n} d_{ij} |\Lambda_{m+i}^{(j)} \rangle $\end{tabular}                   & \begin{tabular}[c]{@{}l@{}}$W_{N_{R_{ij}};I_{p}^{(l)} I_{m+q}^{(h)}} 
= (\delta_{ip} \delta_{jl} \delta_{q1}\delta_{h0} $\\$+
\delta_{iq} \delta_{jh} \delta_{p1}\delta_{l0}) \theta_2$\end{tabular}      & \begin{tabular}[c]{@{}l@{}}$C+D=\frac{1}{\omega \lambda}\left(\rho^{(R;-2)}_{N_{R_{ij}},0_R}\right)_{m \times n}$ \\~\\ $\omega=  \theta_2 c_{00}^* d_{00}^*$\end{tabular}          \\ \hline

Determinant of matrix \\ $E=(e_{ij})_{n \times n}$ & \begin{tabular}[c]{@{}l@{}}$S_i:~|\psi_i \rangle=e_{i0}|0\rangle$\\$+\sum_{j=1}^{n}e_{ij}| \Phi_{i}^{(j)} \rangle$\end{tabular}                   & $W_{1_R;I_1^{(i_1)}\cdots I_{n}^{(i_n)}}=\epsilon_{i_1i_2\cdots i_n}\theta_3$   & \begin{tabular}[c]{@{}l@{}}$\det(E)=\frac{1}{\gamma} \;\rho_{1_R;0_R}^{(R;-n)}$ \\~\\ $\gamma=\theta_3\prod\limits_{i = 1}^n {{e_{i0}}^*}$\end{tabular} \\ \hline

Inverse of matrix  \\ $E=(e_{ij})_{n \times n}$   & \begin{tabular}[c]{@{}l@{}}$S_i:~|\psi_i \rangle=\hat e_{i0}|0\rangle+\sigma| \Xi_{i}^{(n+1)} \rangle$\\$+\sum_{j=1}^{n}e_{ij}| \Xi_{i}^{(j)}\rangle$ \end{tabular}                   & \begin{tabular}[c]{@{}l@{}}$W_{\hat N_{R_{ij}};I_1^{(l_1)}\cdots I_i^{({n+1})}\cdots I_{n}^{(l_n)}}$ \\ 
$=(-1)^{i+j} \epsilon_{j;l_1\cdots l_{i-1}l_{i+1}\cdots l_n}\theta_4$\\ (for complement) \\~\\ \hline \\ $W_{\hat N_R;I_1^{(i_1)}\cdots I_{n}^{(i_n)}}=\epsilon_{i_1i_2\cdots i_n}\theta_3$ \\(for determinant)\end{tabular}     & \begin{tabular}[c]{@{}l@{}l@{}}$E_{ij}=\frac{1}{\mu}~ \rho_{\hat N_{R_{ij}};0_R}^{(R;-n)}$ \\~\\ $\mu=\theta_4\sigma\prod\limits_{i = 1}^n {{\hat e_{i0}^*}}$ \\~\\ \hline \\ $\det(E)=\frac{1}{\hat \gamma} \;\rho_{\hat N_{R};0_{R}}^{(R;-n)}$ \\~\\ $\hat \gamma=\theta_3\prod\limits_{i = 1}^n {{\hat e_{i0}^*}}$ \\~\\ \hline \\ $E^{-1}(j,i)=\frac{\rho_{\hat N_{R_{ij}};0_R}^{(R;-n)}}{\sigma\sqrt{n}\rho_{\hat N_{R};0_{R}}^{(R;-n)}}$\end{tabular}       \\~\\ \hline
\end{tabular} 
\caption{Summary of quantum algorithms for matrix operations. 
Elements of the concerned matrices are encoded as the {\it probability amplitudes} in the pure normalized states $|\psi_i\rangle$. The receiver's state, in the multi-index notation, is written as 
$\rho^{(R;-n)}_{N_R;M_R} =\sum_{N^{\prime},I,J} W_{N^{\prime}N_R;I}
~\rho _{I_{S_1};J_{S_1}}^{(S_1)} \cdots
\rho_{I_{S_n};J_{S_n}}^{(S_n)}~W_{J;N^{\prime}M_R}^{\dagger}$, where $X \equiv X_1 \cdots X_r$ ($X = N^{\prime},~I,~J$), $I_{S_i} \equiv I_1 \cdots I_{r_i}$, and so on.
See text for details. 
Here $\theta_1 = \frac{1}{\sqrt{k}}$, $\theta_2 = \frac{1}{\sqrt{2}}$, $\theta_3 = \frac{1}{\sqrt{n!}}$, and $\theta_4 = \frac{1}{\sqrt{(n-1)!}}$. 
For every matrix operation, the unitary transformation $W$ is universal, i.e., it does not depend on particular matrices encoded into the senders' states.}
\label{table:quant-algo}
\end{table*}

\section{Conclusion}\label{sec:conclusion}
Based on the ``Sender-Receiver" model, we have constructed the quantum computational framework for calculating  matrix-vector product, matrix-matrix product, sum of two matrices, and determinant and inverse of a matrix over complex field. By encoding the information of matrices into the pure initial states of senders and performing the appropriate unitary transformation $W$, the final results are elements of the receiver's density matrix which, in turn, can be considered as input for other quantum algorithms. 
For every matrix operation, the unitary transformation $W$ is universal, i.e., it is independent of particular matrices encoded into the senders’ states.
Finally, as an application of the proposed quantum algorithms, in particular, inverse of a matrix, we have presented a quantum scheme for solving linear systems of equations different from the HHL algorithm. The proposed algorithms also provide new insights to develop other quantum algorithms.

\begin{acknowledgments}
This project is supported by the National Natural Science Foundation of China (Grant No. 12031004 and Grant No. 61877054), and the Fundamental Research Foundation for the Central Universities (Project No. K20210337). The work was partially funded by a state task of Russian Fundamental Investigations (State Registration No. AAAA-A19-119071190017-7).
\end{acknowledgments}


\begin{thebibliography}{10}
\bibitem{note-running-time} 
A $d\times d $ unitary transformation can be represented as a product of $\frac{d(d-1)}{2}$ two-level unitary gates~\cite{QFT3}. In our algorithms, we use $k$-excitation blocks $W_k$ of a unitary transformation $W$ of an $N$-qubit system, where $N$ and $k$ depend on a particular algorithm, and the dimension of $W_k$ is $d=\binom{N}{k}$.
In the algorithms of matrix-vector product, matrix-matrix product, and sum of matrices, $k=2$ and $\frac{d(d-1)}{2}=O(N^4)$. Hence, the unitary operation $W_2$ can be implemented using a circuit containing $O(N^6)$ single qubit and CNOT gates.
\bibitem{note0}
If a matrix $M$ does not satisfy the condition $\left\| M \right\|_2^2 < 1$, it can be multiplied by a small enough number on a classical computer so that the Frobenius norm of the resulting matrix  is less than one. Without loss of generality, all matrices are constrained by this requirement.
\bibitem{note1} The density matrix of an $N$-spin quantum system can be written as a sum~\cite{EBF} $\rho = \sum_{n={-N}}^N  \rho^{(n)}$,
where each submatrix $ \rho^{(n)}$  consists of the elements of $\rho$ responsible for spin-state transitions that change the number of excited spins by $n$. This number is positive, negative or zero respectively when the number of excited spins increases, decreases or remains the same after such transition. Matrices $\rho^{(n)}$ are called  multi-quantum coherence matrices and the number $n$ is called the order of coherence matrices.
The above representation is especially evident if we go to the basis of $N$-qubit quantum states sorted by the excitation number.
\end{thebibliography}
\end{document}